# Hybrid superconductor-semiconductor devices made from self-assembled SiGe nanocrystals on silicon

G. Katsaros,[1,*] P. Spathis,[1] M. Stoffel,[2] F. Fournel,[3] M. Mongillo,[1] V. Bouchiat,[4] F. Lefloch,[1] A. Rastelli,[2] O. G. Schmidt,[2] and S. De Franceschi [1,*]

The epitaxial growth of germanium on silicon leads to the self-assembly of SiGe nanocrystals via a process that allows the size, composition and position of the nanocrystals to be controlled. This level of control, combined with an inherent compatibility with silicon technology, could prove useful in nanoelectronic applications. Here we report the confinement of holes in quantum-dot devices made by directly contacting individual SiGe nanocrystals with aluminium electrodes, and the production of hybrid superconductor-semiconductor devices, such as resonant supercurrent transistors, when the dot is strongly coupled to the electrodes. Charge transport measurements on weakly coupled quantum dots reveal discrete energy spectra, with the confined hole states displaying anisotropic gyromagnetic factors and strong spin-orbit coupling strength with pronounced gate-voltage and magnetic-field dependence.

[1]CEA, INAC/SPSMS/LaTEQS, 17 Rue des Martyrs, 38054 Grenoble, France

[2]IFW-Dresden, Institute for Integrative Nanosciences, Helmholtzstrasse 20, 01069 Dresden, Germany

[3]CEA, LETI, MINATEC, F38054 Grenoble, France

[4]Institut Néel, CNRS and Université Joseph Fourier, BP 166, 38042 Grenoble cedex 9, France

*To whom correspondence should be addressed: georgios.katsaros@cea.fr, silvano.defranceschi@cea.fr.

The bottom-up fabrication of electronic devices from predefined nanoscale structures is a major theme in nanoelectronics, and different types of nanostructures have been considered as potential building blocks for such devices. Carbon nanotubes and semiconductor nanowires, for example, have been used to fabricate high-performance field-effect transistors and basic functional circuits[1-4]. While these achievements represent important steps towards low-cost electronics, there are challenges to overcome with respect to integration with mainstream silicon technology and scaling towards high-density electronic circuits with large numbers of interconnected devices.

Here we propose a new approach to bottom-up nanodevices based on self-assembled silicon-germanium (SiGe) nanocrystals grown directly on Si by molecular-beam epitaxy via the so-called Stranski-Krastanow growth mode[5,6]. These nanocrystals can have a variety of sizes and shapes (Fig. 1a)[7,8], and their composition can be controlled to a high degree by adjusting the growth parameters[9]. In addition, their positions can be controlled via prepatterning of the growth surface as shown by the example in the lower part of Fig. 1a [10,11]. However, despite being potentially scalable and compatible with CMOS technology[12], there have been no reports of practical devices based on self-assembled SiGe nanocrystals to date. Another advantage of SiGe nanostructures is their ability to form ideal contacts with metals, which is essential for making hybrid superconductor-semiconductor devices, as demonstrated in the recent observation of a gate-tunable supercurrent (of holes) in Ge/Si core/shell nanowires[13]. Here we exploit the smaller contact area and the lower dimensionality of self-assembled SiGe nanocrystals to extend gate control of the supercurrent to the single-hole level.

SiGe nanostructures could also have applications in semiconductor spintronics because of their potential for long spin coherence times[14-18]. While both n- and p-type structures have been recently addressed, the latter bear some additional interest mainly because of the stronger spin-orbit (SO) coupling of valence-band states. SO coupling can induce large modulations and anisotropies in the Landé g-factor[19-22] enabling electrically-controlled spin precession[23]. So far experimental efforts have focused on coupled quantum-dot systems (created by local electrostatic gating in p-type Ge/Si core/shell nanowires)[14], and indirect evidence for SO coupling has been found in the g-factor of weakly confined hole states[24]. The quantum-dot devices made from self-assembled SiGe nanocrystals demonstrated here show more pronounced SO effects in the strong quantum confinement regime, and spin-1/2 hole states with largely anisotropic g-factors when the quantum dot contains and odd number of holes. In addition to spintronics, self-assembled SiGe nanocrystals may facilitate the development of two-dimensional qubit architectures in quantum information applications, as opposed to one-dimensional architectures based on Ge/Si core/shell nanowires

Inspired by recent work on self-assembled InAs nanocrystals[25-27], we have developed a nanolithographic process to laterally contact pairs of 20-nm-thick aluminium electrodes (to be used as source and drain electrodes) to single self-assembled SiGe nanocrystals. Although SiGe and InAs nanocrystals have similar shapes and sizes, InAs nanocrystals are grown on GaAs rather than Si and, when connected to metal electrodes, they confine electrons rather than holes. Devices made from InAs nanocrystals can be gated by growing the nanocrystals on heavily-doped GaAs-based heterostructures[25-27]. Here, in order to add gate control, the SiGe nanocrystals were grown on silicon-on-insulator (SOI) substrates consisting of an undoped Si overlayer, a SiO2 insulating layer, and a degenerately doped Si substrate that was used as a back gate (see supplementary information). We focused on SiGe nanocrystals that had a characteristic dome-like shape, with a height of 20 and a base diameter of 80 nm, a 2-nm-thick Si capping layer, and a band structure that confines holes (Fig. 1b). At room temperature, the upper Si layer of the SOI substrate introduces a significant parallel conduction path. Below ~100 K, however, transport occurs uniquely by holes tunnelling from the source to the drain via the SiGe quantum dot. All of the measurements on an ensemble of 12 similar devices were taken at 15 mK. We find characteristic device resistances between ~$10^4$ and ~$10^5$ Ohms. The lowest values are close to the resistance quantum, h/e = 25.8 kΩ, denoting high contact transparency.

**Single-hole supercurrent transistor**

At 15 mK the aluminium electrodes are superconducting but they can be turned into normal-type electrodes by applying a perpendicular magnetic field, $B_\perp$, of a few tens mT. The differential conductance, $dI_{SD}/dV_{SD}$, of a low-resistance device is shown in Fig. 2a as a function of back-gate voltage, $V_G$, at $B_\perp$ = 75 mT. The observed oscillations are a consequence of the on-site Coulomb interaction forcing holes to tunnel one by one across the SiGe quantum dot. The conductance valleys correspond to the Coulomb blockade regime wherein the quantum dot hosts an integer number of confined holes. Each conductance peak corresponds to an energy degeneracy between consecutive charge states. The large width of the Coulomb peaks and the finite valley conductance denote a strong tunnel coupling to the source and drain leads. As the contact electrodes are turned into a superconducting state by removing the magnetic field, this strong coupling enables the onset of Cooper-pair tunneling across the quantum dot leading to measurable supercurrents[28-33]. This non-dissipative transport mechanism is modulated by the gate voltage as it can be seen from the current-biased measurement shown in Fig. 2b. The measured voltage-drop across the SiGe quantum dot, shown in colour scale, is reducing inside the black regions around the charge degeneracy points indicating resonant supercurrent transport. A representative $V_{SD}(I_{SD})$ trace taken at one of such Coulomb-blockade resonances (blue dashed line in Fig. 2b) is given in Fig. 2c (blue trace). The device switches from superconducting to dissipative regime at a bias current of

~$10^2$ pA. No supercurrent branch is observed in the adjacent Coulomb valley (green dashed line in Fig. 2b) as shown by the representative green trace in Fig. 2c. As a result, at low current bias the device can be turned from a superconducting "ON" state to a dissipative "OFF" state by a small change in the gate voltage corresponding to a fractional variation of the device charge. This operating principle is illustrated in Fig. 2d.

**Tunable hole spin states**

We now consider the opposite case of a high-resistance device and focus on the spin-dependent properties of self-assembled SiGe quantum dots. In the normal state ($B_\perp$ = 50 mT), well-separated Coulomb blockade resonances are observed in a measurement of $I_{SD}$ vs $V_G$ (Fig. 3a), with the device conductance vanishing in the Coulomb valleys. Additional information is obtained by plotting $I_{SD}$ vs ($V_G, V_{SD}$) as shown in Fig. 3b. The Coulomb blockade regime takes place within the regions with the characteristic diamond shape. On average, the diamond size grows from left to right denoting an increase of the charging energy from ~5 to ~20 meV which follows mainly from a decrease in the tunnel and capacitive coupling between the quantum dot and the leads.

By zooming into the rightmost charge degeneracy point in Fig. 3b, additional features become visible (Fig. 3c). At zero field, the Coulomb diamonds are slightly split apart along the $V_{SD}$ axis leading to the appearance of a "currentless window" around zero bias. This feature is due to the superconducting gap in the quasiparticle density of states of the electrodes (Cooper pair tunneling and sub-gap transport are suppressed for high tunnel resistances). A second important feature is the presence of additional current steps appearing as lines parallel to the diamond edges (see yellow arrows in Fig. 3c). These steps arise from single-hole tunnelling via higher-energy orbital levels demonstrating that SiGe nanocrystals form true quantum dots with discrete energy spectra.

At finite perpendicular fields, the Zeeman spin splitting of the discrete quantum dot levels is revealed by the appearance of new excitation lines such as those indicated by black arrows in Fig. 3c. A similar behaviour is seen for fields parallel to the substrate. The observed two-fold splitting of the right-diamond edges denotes a ground state with spin S = 1/2 in the left-diamond. The absolute value, g, of the hole g-factor can be extracted from the magnitude of the Zeeman splitting, $\Delta E_Z = g\, \mu_B\, B$, where $\mu_B$ is the Bohr magneton. The procedure is illustrated in Fig. 3c and in Figs. 3d,e. The measured g-factors differ substantially from the free-electron value (2.002) and exhibit a pronounced anisotropy, with $g_\perp$ =2.71 and $g_{//}$ =1.21 being the perpendicular- and parallel-field values, respectively. This anisotropy is qualitatively consistent with recent calculations for pure Ge islands with pyramidal shape[34]. Similar anisotropies have been reported also for strained bulk Ge[35], acceptor levels in Si/Ge/Si heterostructures[36], and Ge/Si core/shell nanowires[24].

To investigate the dependence of the g-factor on the number of confined holes, we have carried out similar measurements for other charge degeneracy points in the $V_G$ range of Fig. 3b. We find an alternation of S=0 and S=1/2 ground states corresponding to an even and odd filling of spin-degenerate levels, respectively. As noticed above, lowering $V_G$ leads to a larger tunnel coupling between the SiGe quantum dot and the metal contacts, resulting in a larger energy broadening, Γ, of the quantum dot levels. Since the resolution of single-hole tunneling spectroscopy is limited by Γ, the experimental uncertainty on $\Delta E_Z$ increases. At sufficiently large tunnel coupling, however, two-electron processes begin to contribute a measureable current in the Coulomb blockade regime providing a powerful spectroscopy tool. In fact cotunneling processes can induce internal excitations at finite bias. The onset of these so-called "inelastic" processes occurs when $eV_{SD}$ equals the energy to create an excitation in the quantum dot, leading to a step-like increase in $dI_{SD}/dV_{SD}$ (see Fig. 4a). The step width is uniquely determined by the electronic temperature resulting in a high resolution at low temperatures[37].

This is clearly seen in Fig. 4b where $dI_{SD}/dV_{SD}$ is plotted as a function of ($V_G$,$V_{SD}$) around a charge degeneracy point and for $B_{//}$ = 8 T. The Zeeman splitting is simultaneously visible as a $dI_{SD}/dV_{SD}$ peak in single-hole tunneling and a clearly sharper $dI_{SD}/dV_{SD}$ step in inelastic cotunneling. As expected, the corresponding features merge at the diamond edge. The detailed field-dependence of the Zeeman splitting can be investigated by letting $B_{//}$ (or $B_{\perp}$) vary at fixed $V_G$ inside the Coulomb diamond for a spin-1/2 ground state. One of such measurements is shown in Fig. 4c. The structure around B=0 arises from the superconductivity of the contacts. The stronger tunnel coupling enables in this case sub-gap transport based on the Andreev reflection phenomenon[38]. Above the critical field (650 and 50 mT for parallel and perpendicular fields, respectively) the spin-flip cotunneling steps at $eV_{SD}$ = ± $\Delta E_Z$ shift apart with the applied field.

To determine the full angle dependence of the g-factor, we have varied the field direction while keeping the magnitude constant at 3 T (Fig. 4d). Interestingly, the observed anisotropy does not exactly correspond to the crystal symmetry. The minimum g-factor is offset by 15-20 degrees with respect to the parallel direction. An almost identical offset is found in another diamond for a different number of holes. The same type of data for a different device, however, shows no offset at all. We argue that the observed device-dependence of the offset may originate from a generally asymmetric overlap of the metal contacts with the SiGe nanocrystal and a consequent asymmetry in the confinement potential. In principle, this asymmetry could be reduced by a controlled positioning of the contact electrodes on the SiGe nanocrystal.

Figure 4e provides an overall summary of the g-factor data obtained for five different gate voltages. Both $g_\perp$ and $g_{//}$ as well as their $g_\perp/g_{//}$ ratio exhibit significant variations with the number of confined holes. In fact the g-factor depends on the mixing of heavy-hole and light-hole components, which is expected to change from level to level[34] (see supplementary information). In addition, as the number of holes increases the wave functions of the progressively occupied levels extend more and more into the Si-rich base of the self-assembled nanocrystal. This should lead to an average decrease of the hole g-factor in line with our experimental finding.

G-factor differences can also be observed between ground-state and excited-state levels measured in the same charge regime, i.e. at approximately the same $V_G$ (Fig. 3e). On the other hand, the g-factors are found to be rather insensitive to $V_G$ variations within the same Coulomb diamond. We conclude that the g-factors are clearly linked to the corresponding orbital wave functions and that the latter appear to be only weakly affected by gate variations corresponding to the full width of a Coulomb diamond. Alternative gate geometries (e.g. dual-gate devices) may possibly result in a more efficient g-factor tuning at constant number of holes.

A summary of the g-factors measured at different charge numbers is given in Fig. 4e. The reported values are obtained from a linear fit of the data in the high-field regime. In fact, an appreciable nonlinearity is found in the relation between $\Delta E_Z$ and $B_{//}$ (or $B_\perp$) which can be seen as a field-dependent g-factor. In some cases, the g-factor can increase by as much as ~75% for magnetic fields in the experimentally accessible range (see Fig. 4f).

**Measurement and anisotropy of the spin-orbit coupling strength**

The observed g-factor deviations from the free-particle value and the nonlinearities in the Zeeman effect constitute indirect evidence of a strong coupling between orbital and spin degrees of freedom. Although this coupling can cause spin relaxation[39] via the interaction with phonons, it can as well provide a useful handle for coherent spin manipulation by means of gate-controlled electric fields[40-43]. In order to gain direct quantitative information on the SO coupling strength, we considered devices with relatively small level spacing and, incidentally, stronger tunnel couplings. A plot of $dI_{SD}/dV_{SD}$ vs ($V_G,V_{SD}$) is shown in Fig. 5a for one of such devices at $B_\perp$ = 2 T. Multiple $dI_{SD}/dV_{SD}$ steps can be seen in two adjacent diamonds denoting the contribution of different orbital levels to the inelastic-cotunneling current. To identify the precise origin of these features, we fixed $V_G$ in correspondence of the blue line in Fig. 5a and let $B_{//}$ vary between 0 and 6 T. The result is shown in Fig. 5b where $dI_{SD}/dV_{SD}$ has been replaced by its numerical derivative $d^2I_{SD}/dV_{SD}^2$ to emphasise the onset of inelastic cotunneling transitions.

The superconductivity-related structure around $B_{//}$ = 0 is discussed in the supplementary information. Above the critical field, we identify three departing lines which can be ascribed to the excitations from a spin-singlet ground state, |0,0⟩, to three spin-triplet excited states, denoted as |1,+1⟩, |1,0⟩, and |1,−1⟩. This assignment implies an even number of confined holes. The corresponding energy diagram is qualitatively shown in Fig. 5d. The zero-field singlet-triplet splitting is ~130 μeV. At about 2 T an anticrossing is observed between the field-independent |0,0⟩ state and the |1,1⟩ state. This anticrossing between normally orthogonal states indicates the existence of mixing via SO coupling. An estimate of the coupling strength, $\Delta_{SO}$, can be directly extracted from the minimum level splitting (2$\Delta_{SO}$). We find $\Delta_{SO}$ = 34 μeV, which is less than an order of magnitude smaller than in InAs[44] or InSb[22] nanowires. The vertical arrows indicate the possible transitions resulting from inelastic cotunneling. The transition from |1,+1⟩ to |1, −1⟩, denoted by a dashed blue arrow in Fig. 5d,

would be prohibited in the absence of SO coupling since it requires a change in $S_z$ larger than 1 [45]. For systems with strong SO coupling, however, states are no longer pure singlet or triplet and thus such a transition becomes possible. A similar behaviour is observed for perpendicular fields (see Fig. 5c). Yet we find $\Delta_{SO}$ = 42 μeV denoting a dependence of the SO coupling strength on the field direction.

To further investigate this effect, we carried out the same study in the next diamond, corresponding to an odd number of holes. The observed inelastic cotunneling steps can be ascribed to the splitting of two subsequent orbital levels with a zero-field energy difference of ~300 μeV. The data, taken along the red line in Fig. 5a are shown in Figs. 5e and 5f for parallel and perpendicular fields, respectively. The corresponding qualitative energy diagrams are given in Figs. 5g and 5h, where we have indexed the spin states with the quantum numbers, $n$ and $n+1$, of the corresponding orbitals. To obtain the best qualitative matching, we have assumed a field-induced decrease in the orbital splitting and slightly different g-factors for the two subsequent orbital levels.

We observe an anticrossing between $|\downarrow\rangle_n$ and $|\uparrow\rangle_{n+1}$ for $B_{//}$~2.6 T (red disk in Fig. 5g) corresponding to $\Delta_{SO}$ = 37 μeV. This anticrossing turns into a crossing at $B_\perp$~1.5 T denoting vanishing SO coupling strength. This result is the most striking manifestation of the interplay between SO coupling and an external magnetic field, an effect that was recently predicted in a theoretical work by Golovach *et al.* for the case of GaAs quantum dots[46]. Interestingly, the observed punctual suppression of the SO coupling strength should result in a longer spin relaxation time. (During the preparation of this manuscript we have become aware of an experimental work in which a similar anisotropy of the SO coupling strength was observed for electrons in InAs quantum dots[47].) The other anticrossings in Figs. 5e and 5f cannot be used for an estimate of $\Delta_{SO}$ since they occur between levels that could anticross even in the absence of SO coupling due to purely orbital mixing.

**Summary and outlook**

The measurements of single-hole tunnelling and two-hole co-tunnelling presented here provide fresh insights into the electronic properties of self-assembled SiGe nanocrystals. We have observed finite-size quantum confinement and various effects associated with a strong and tunable SO coupling. We have also shown that it is possible to form low-resistance contacts to superconducting electrodes, and thus demonstrated the first example of a single-hole supercurrent transistor based on SiGe. In addition to potential device applications, self-assembled SiGe nanocrystals also provide a new versatile playground for investigating a variety of quantum phenomena in condensed matter physics. In particular, access to the strong-coupling limit could open up new opportunities to explore spin-orbit physics and other spin-related phenomena, such as the Kondo effect in combination with superconducting and possibly ferromagnetic correlations[26,27,48-50].

**Acknowledgments**

The authors acknowledge T. Haccart and the PTA cleanroom team of CEA, J.-L. Thomassin and F. Gustavo for their help in device fabrication, and T. Fournier for helpful discussions and free access to fabrication recipes and equipment at the NANOFAB facility of the Néel Institute. The authors acknowledge helpful discussions with M. Houzet, V. Golovach, W. Wernsdorfer, D. Feinberg, G. Usaj, R. Whitney, M. Sanquer, X. Jehl, G. A. Steele and E. J. H. Lee. Supported by the Agence Nationale de la Recherche through ACCESS and COHESION projects. G.K. acknowledges further support from the Deutsche Forschungsgemeinschaft (grant KA 2922/1-1).

**Author contribution statement:**

G. K and S. D. F. planned the experiment, interpreted the data and co-wrote the paper. G.K. fabricated the devices, performed together with P.S. and S. D. F. the measurements and analyzed the data. P. S. participated in the analysis of the data and set up the dilution

**Additional Information**

Supplementary information accompanies this paper at www.nature.com/naturenanotechnology. Reprints and permission information is available online at http://npg.nature.com/reprintsandpermissions/. Correspondence and requests for materials should be addressed to G.K. or S.D.F.

**Figure 1. Structure and growth of the SiGe self-assembled nanocrystals and device layout**

**a,** Top: Three-dimensional scanning-tunnelling micrographs of self-assembled SiGe nanocrystals with characteristic 'hut', 'pyramid', and 'dome' shapes. The corresponding dimensions are 50x32x7 nm$^3$, 50x32x7 nm$^3$ and 50x50x10 nm$^3$, respectively. These sizes can be tuned by adjusting growth conditions. For the present work we used growth conditions yielding dome-shaped monocrystals with a height of ~20 nm and a base diameter of ~80 nm. Bottom: Atomic-force micrograph (4.7x4.7 μm$^2$) illustrating an example of a self-organized array of SiGe nanocrystals grown on a prepatterned Si wafer. **b,** Schematic of a quantum-dot device obtained by contacting a single SiGe nanocrystal to aluminium source/drain electrodes. The heavily doped substrate is used as a back gate. Top-right panel: Scanning-electron micrograph of a representative device (scale bar: 100 nm). Top-left panel: Schematic cross-sectional view of a device and corresponding qualitative band diagram with valence- and conduction-band profiles shown as green and red lines, respectively. The SiGe nanocrystal, which is covered by a 2-nm-thick Si layer, acts as a confining potential for holes (quantized level schematically shown as a set of black horizontal lines). The valence-band edge lies close to the Fermi energies, $\mu_S$ and $\mu_D$, of the source and drain electrodes. This band alignment is consistent with the one given in Ref. 2 for Ge/Si core/shell nanowires.

**Figure 2. SiGe single-hole supercurrent transistor**

**a,** Zero-bias differential conductance, $dI_{SD}/dV_{SD}$, versus back-gate voltage, $V_G$, showing Coulomb-blockade oscillations in a low-impedance device at 15 mK and with a 75 mT perpendicular magnetic field suppressing superconductivity in the Al electrodes. **b,** Differential resistance, $dV_{SD}/dI_{SD}$, on colour scale for the same $V_G$ range at zero magnetic field, i.e. with superconducting electrodes. Resonant supercurrents can be clearly observed at the position of the charge degeneracy points as black regions. The bright regions in between denote the Coulomb blockade regime. **c,** Representative $V_{SD}(I_{SD})$ traces extracted from **(b)** at the blue and green vertical lines illustrating the device behaviour on (blue) and off (green) resonance. A series resistance of about 40 kΩ corresponding to low-temperature low-pass filters, wiring, and measurement electronics has been substracted in both **(b)** and **(c)**. **d,** Qualitative electronic density of states (horizontal axis) versus energy (vertical axis) for a SiGe quantum dot between Al superconducting leads. In the leads, an energy gap Δ separates the condensate of Cooper pairs at the Fermi energy from occupied and unoccupied single-particle states. In the quantum dot, the discrete hole levels are shown as peaks with a life-time broadening due to tunneling. The gate voltage tunes the energy of the quantum-dot levels with respect to the Fermi energy of the leads. Since Cooper-pair tunnelling takes place on resonance (blue line) and it is suppressed off resonance (green dashed line), the device can be electrically switched from a superconducting to a dissipative state by a small change in $V_G$.

**Figure 3. Tunneling spectroscopy measurements on a high-resistance device**

**a,** Representative Coulomb-blockade oscillations in $I_{SD}(V_G)$ at 15 mK and $V_{SD}$ = 1 mV. **b,** $I_{SD}(V_G,V_{SD})$ on colour scale at a 50-mT perpendicular magnetic field suppressing superconductivity in the contacts. **c,** Colour plots of $I_{SD}(V_G,V_{SD})$ at the dotted rectangle in **(b)** for different parallel and perpendicular magnetic fields. $V_{SD}$ spans an 8 mV range around zero bias, $V_G$ a 100 mV range around the charge degeneracy point. The yellow arrows indicate the onset of single-hole tunneling via an excited orbital state. Finite parallel or perpendicular magnetic fields cause the edges of the right Coulomb diamond to split due to lifted degeneracy in the spin-1/2 ground state of the left diamond. The black arrows indicate the onset of single-hole tunneling via the spin-down ($|\downarrow\rangle$) excited state as represented by the energy diagram in the inset. (Because SiGe nanocrystals consist mainly of Ge, hole g-factors are likely negative. Hence $|\uparrow\rangle$ and $|\downarrow\rangle$ correspond to spin parallel and antiparallel to the applied field, respectively.) The Zeeman splitting, $\Delta E_Z$, between $|\uparrow\rangle$ and $|\downarrow\rangle$ states can be extracted from the splitting of the corresponding diamond edges. **d,** $\Delta E_Z$ versus magnetic field (parallel or perpendicular) as extracted from **(c)** and similar measurements. For each field, $\Delta E_Z$ is obtained after averaging on different $V_G$ values and error bars are determined by the resulting standard deviation. Solid lines are fits to a linear dependence $\Delta E_Z = g\mu_B B$. Anisotropy between in-plane and perpendicular g-factors is observed. **e,** $dI_{SD}/dV_{SD}(V_G,V_{SD})$ for an 8-T parallel field. The Zeeman splitting is observed for ground and excited orbital states, yielding $\Delta E_Z$ and $\Delta E^*_Z$, respectively.

**Figure 4. Anisotropy and gate dependence of the hole g-factors**

**a,** Schematic energy diagram showing the onset condition ($eV_{SD} = \Delta E_Z$) for spin-flip inelastic cotunneling. **b,** $dI_{SD}/dV_{SD}(V_G,V_{SD})$ for a parallel magnetic field of 8 T. The Zeeman splitting ($\Delta E_Z$) of the spin-1/2 ground state in the left Coulomb diamond can be accurately measured from the $V_{SD}$ position of the $dI_{SD}/dV_{SD}$ step due to inelastic spin-flip cotunneling. **c,** Magnetic-field evolution of this inelastic cotunneling step at fixed $V_G = 3.52$ V. **d,** Angle dependence of the spin-flip inelastic cotunneling edges at a fixed $V_G$ (3.1 V) and magnetic-field amplitude (3T). The position of the minimum splitting does not correspond to the parallel direction as indicated by the white arrow. As a result, the minimum g-factor, $g_{min} \sim 0.5$, is significantly smaller than the zero-angle g-factor, $g_{//} \sim 0.8$, and the maximum anisotropy, $g_{max}/g_{min} \sim 5$, is consequently larger than $g_\perp/g_{//} \sim 3.1$. **e,** Summary of the parallel and perpendicular g-factors (absolute values) measured on the same device at different gate voltages. For a better comparison between the g-factor values extracted from direct-tunneling data and those from cotunneling data, linear fits were always taken in the large magnetic-field range. Inset: $V_G$–dependence of the g-factor anisotropy. **f,** Evolution of the Zeeman splitting as measured from the inelastic cotunneling spectroscopy at a fixed $V_G = 3.1$ V for parallel and perpendicular magnetic fields. The dashed lines are linear fits at high magnetic fields and the extracted values are the ones reported in the Fig. 4e. Over the entire range of parallel and perpendicular fields, however, the g-factors exhibit an appreciable nonlinearity which can be fitted to a power-law dependence. We obtain $\Delta E_Z \sim 2.06\mu_B B_\perp^{1.2}$ and $\Delta E_Z \sim 0.71\mu_B B_{//}^{1.07}$ for perpendicular and parallel magnetic fields, respectively.

**Figure 5. Anisotropic spin-orbit coupling strength probed by inelastic cotunneling**

**a,** $dI_{SD}/dV_{SD}(V_G,V_{SD})$ at $B_{//} = 2$ T. **b-c,** $d^2I_{SD}/dV_{SD}^2$ versus ($B_{//},V_{SD}$) or ($B_\perp, V_{SD}$) for fixed $V_G$ at the blue line in **(a)** corresponding to an even number of confined holes. The vertical arrows in **(b)** indicate inelastic cotunneling transitions for positive $V_{SD}$. **d,** Qualitative energy diagram accounting for the data in **(b-c)**. The spin-triplet excited state exhibits a three-fold splitting in either parallel or perpendicular magnetic fields. Due to spin-orbit coupling, the lowest energy triplet component, $|1,-1\rangle$, and the spin-singlet ground state, $|0,0\rangle$, anticross each other at $B_{//}$ ~ 2 T and $B_\perp$ ~ 1.5 T. The cotunneling transition from $|1,-1\rangle$ to $|1,1\rangle$ (dashed vertical arrow) would be forbidden in the absence of spin-orbit coupling. **e-f,** $d^2I_{SD}/dV_{SD}^2$ versus $(B_{//}, V_{SD})$ or $(B_\perp, V_{SD})$ for fixed $V_G$ at the red line in **(a)** corresponding to an odd number of confined holes. **g-h**, Qualitative energy diagrams illustrating the splitting of two subsequent orbital levels, as observed in **(e)** and **(f)** for parallel and perpendicular fields, respectively (field ranges match qualitatively those of the corresponding measurements). Interestingly, the anticrossing highlighted by a the red disk, between opposite spin states $|\downarrow\rangle_n$ and $|\uparrow\rangle_{n+1}$, which is observed at $B_{//}$ ~ 2.6 T in **(e),** turns into a crossing when the field is applied in the perpendicular direction (red square), as observed at $B_\perp$ ~ 1.5 T in **(f)**. This demonstrates a strong dependence of the spin-orbit coupling strength on the field direction. The other anticrossings in **(f)** and **(e)**, corresponding to the blue and green disks in **(g)** and **(h)**, might be due to purely orbital mixing.

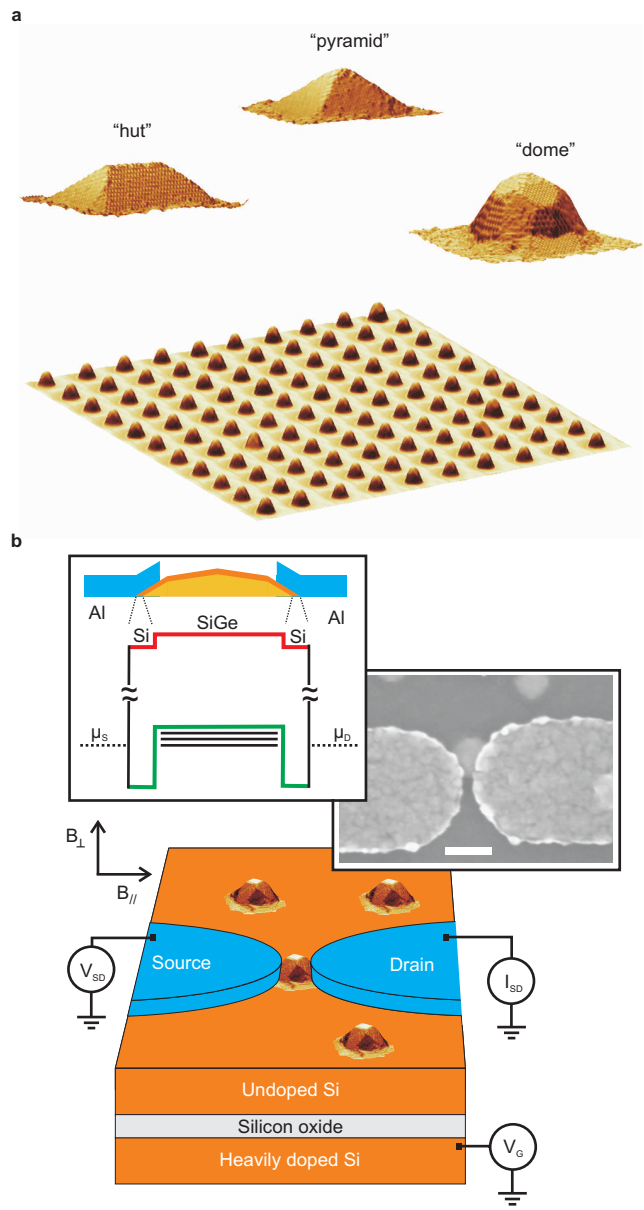

Fig. 1

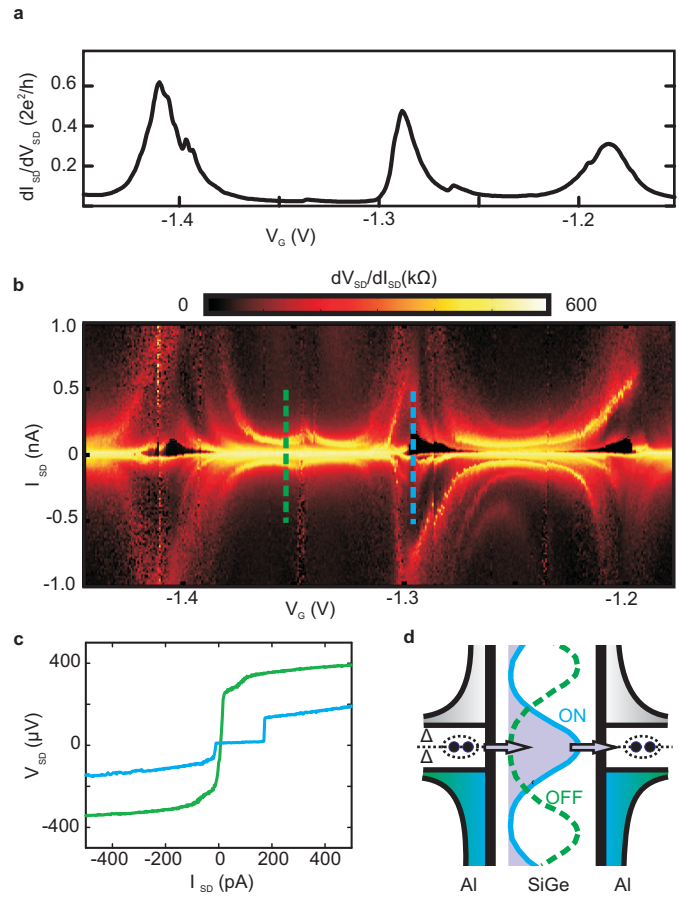

Fig. 2

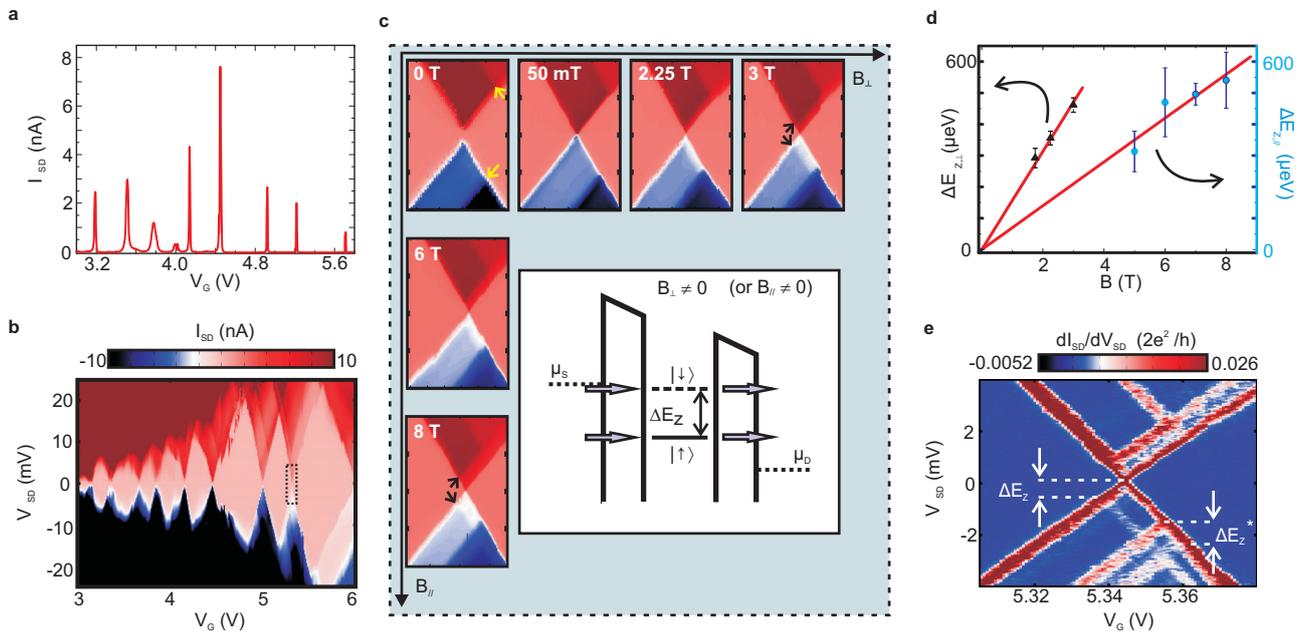

Fig. 3

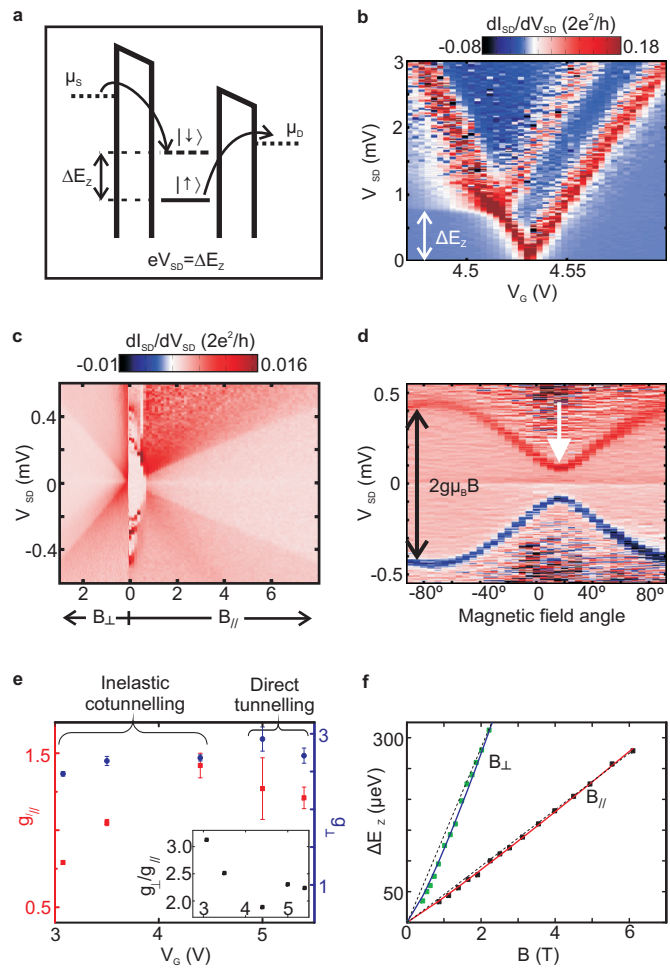

Fig. 4

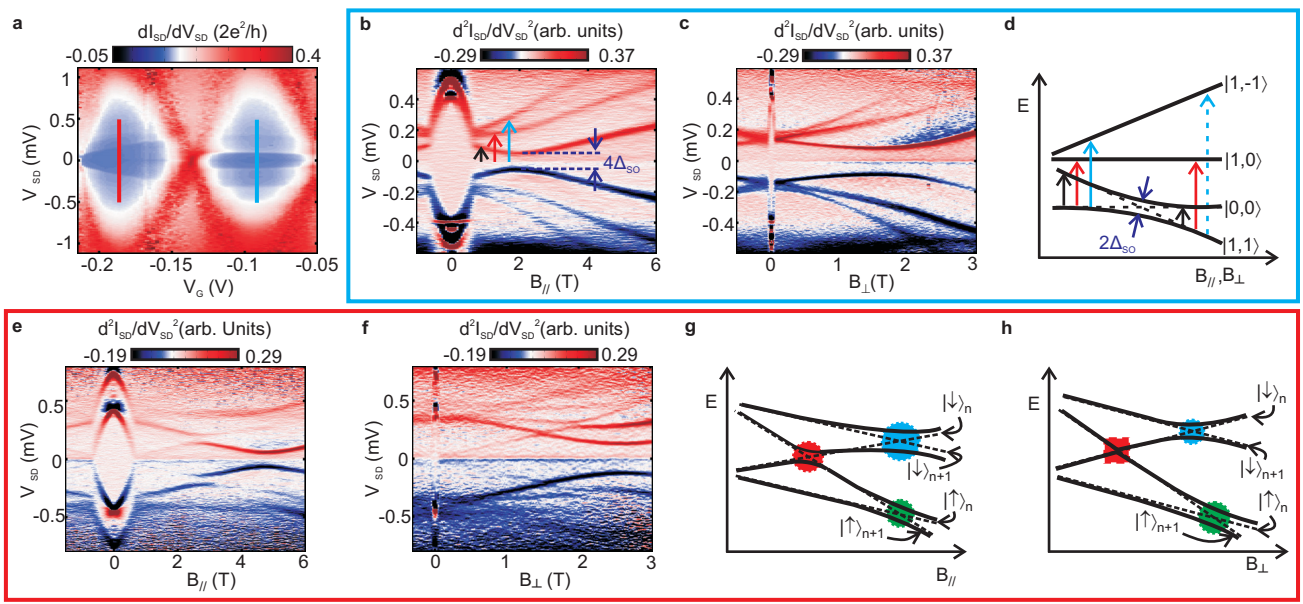

Fig. 5

**Hybrid superconductor-semiconductor devices made from self-assembled SiGe nanocrystals on silicon**

G. Katsaros, P. Spathis, M. Stoffel, F. Fournel, M. Mongillo, V. Bouchiat, F. Lefloch,

A. Rastelli, O. G. Schmidt, S. De Franceschi

This pdf file contains:

Supplementary text

Supplementary figure

References

Supplementary text

1. Fabrication of ad-hoc silicon on insulator (SOI) substrates

Commercial 200-mm-diameter SOI wafers with a Si upper layer of 70 nm, a Si oxide layer of 145 nm (box layer), and a 0.5-mm-thick undoped Si substrate (handle layer), were thermally oxidized and bonded by means of hydrophilic molecular bonding to a heavily doped Si wafer (resistivity of 0.006-0.010 Ohm.cm$^{-1}$). The SOI handle layer was then removed by grinding and selective chemical etching with a 25% diluted solution of tetramethyl ammonium hydroxide (TMAH) at 75°C. Finally, also the box oxide layer of the original SOI wafer was removed using a 10% fluorhydric acid solution. The final result consisted of non standard SOI substrate having 40-nm-thick Si upper layer, a 65-nm-thick oxide layer (obtained by partial thermal oxidation of the Si upper layer in the original SOI wafer) and a degenerately doped Si substrate.

2. Growth of SiGe self-assembled nanocrystals

The SiGe nanocrystals were grown by solid-source molecular beam epitaxy (MBE) on the SOI substrate described above. After ex-situ chemical cleaning, an additional HF dip was performed to remove the surface oxide. The sample was then transferred into the MBE chamber and outgassed at 620°C prior to the deposition of a 100-nm-thick undoped Si buffer at a rate of 0.1 nm/s. After a 5 s growth interruption, 7 monolayers (ML) of Ge were deposited at 620°C at a rate of 0.04 ML/s. The sample was then cooled down to 300°C and capped with a 2-nm-thick Si layer. Within these growth conditions, the island shape is preserved[1,2]. These growth conditions yield randomly arranged SiGe nanocrystals with monocrystalline, dome-shaped structure, and rather homogeneous size. From analysis of atomic force microscopy (AFM) images we have obtained a height of: 22±0.8 nm and a base diameter of 96±2 nm. It is known that AFM images overestimate the base diameter. From the SEM images we have estimated the real value of the base diameter to be about 80 nm.

The ordered SiGe dots shown in Fig. 1a were grown onto a Si(001) wafer on which prior to growth a two-dimensional array of holes has been patterned by e-beam lithography and reactive ion etching using a $CHF_3/O_2$ plasma[3].

3. Device fabrication

InAs self-assembled nanocrystals have been recently used for the realization of electronic devices with either vertical[4-6] or parallel geometry[7-10]. Inspired by the latter works, we have developed a process to fabricate single-hole transistors based on individual SiGe self-assembled nanocrystals.

The fabrication of SiGe quantum-dot devices was accomplished through four steps of e-beam lithography, e-beam metal deposition, and lift-off. In particular, 1) Ti/Au (10/65 nm) bonding pads, 2) an array of NxM pairs of Ti/Au (2/8 nm) linking pads distributed over a 250x250 µm$^2$ area, 3) an array of NxM pairs of Al (20 nm) electrodes partially overlapping the linking pads and forming gaps of 10-50 nm, and 4) Ti/Au (10/65 nm) electrodes connecting selected pairs of linking pads to the outer bonding pads. In step 3, a 10-15 sec dip in buffered HF was performed prior to Al deposition in order to remove the surface native oxide. Between steps 3 and 4, scanning electron microscopy was used to identify the electrode pairs with a single SiGe nanocrystal.

4. Superconductivity-related effects

Bulk aluminium is a conventional superconductor with superconducting critical temperature, $T_C$ = 1.2 K, and superconducting gap $\Delta$ = 180 μeV. These values are known to increase in thin films[11]. Consistent with this trend we find $\Delta$ = 215 μeV at zero magnetic field and T = 15 mK. This value is extracted from the bias-width of the currentless window around zero bias (Fig. 3c). The currentless window is due to the superconducting nature of the Al electrodes, and to the consequent suppression of quasiparticle transport for $V_{SD}$ between -2$\Delta$/e and 2$\Delta$/e, where $\Delta$ is the superconducting energy gap. Following the appearance of this currentless window, all step-like features in the source-drain current are shifted to higher voltages by exactly 2$\Delta$/e. In addition, steps are transformed into asymmetric peaks reflecting the gap-edge singularities in the density of states of the electrodes[12,13].

We consider now the case of cotunnelling transport in the Coulomb blockade regime. In the normal state, elastic cotunnelling processes can take place at any source-drain voltage (i.e., also around zero bias) leading to a featureless background in the differential conductance, $dI_{SD}/dV_{SD}$. In the case of superconducting electrodes, however, these processes are not allowed for $|V_{SD}|$< 2$\Delta$/e, because they require the transfer of a quasiparticle from the fully occupied band of one contact (below gap) to the empty band of the other contact (above gap), as it shown in Fig. S1b. As a result, the onset of elastic cotunnelling gives rise to $dI_{SD}/dV_{SD}$ peaks at ±2$\Delta$/e[14] which reflect the gap-edge singularities in the quasiparticle density of states of the superconducting electrodes. For the same reason, inelastic cotunnelling processes giving rise to $dI_{SD}/dV_{SD}$ steps at ±2$\delta$/e in the normal state ($\delta$ is an excitation energy of the quantum dot) result in the appearance of $dI_{SD}/dV_{SD}$ peaks at ± ($\delta$+2$\Delta$)/e as it is depicted in Fig. S1c. Increasing the magnetic field causes the gradual suppression of the superconducting gap resulting in the observed inward shift of the elastic and inelastic cotunnelling structures (Fig. S1a).

5. g factors in SiGe QDs

In bulk Si, Ge, and Si$_x$Ge$_{1-x}$ compounds the valence-band edge is characterized by a 4-fold degeneracy between the heavy-hole (HH) and light-hole (LH) states at **k** = 0. This degeneracy is removed by quantum confinement or strain. Depending on the strain sign the HH or the LH states can be lower in energy. For compressing strain, which is the case for the self-assembled nanocrystals, HH states are lower in energy. In a magnetic field, HH states display an anisotropic spin splitting, with $g_\perp$ = 6κ and $g_{//}$ ≈ 0 the g factors along the growth axis and in the quantum-well plane, respectively. Here κ is the so-called Luttinger valence-band parameter. On the other hand, the LH states exhibit an opposite anisotropy, with $g_{//}$ = 4κ and $g_\perp$ = 2κ[15]. The κ parameter is a material-dependent property and it can take either positive or negative values. In particular, κ = 0.42 in Si and κ = -3.37 in Ge leading to positive and negative g factors, respectively[16]. In SiGe alloys, the κ parameter takes intermediate values depending on the relative amount of Ge and Si. According to recent calculations[17], κ varies from −0.308 to −1.153 when the Ge content is increased from 60 to 80%; for low Ge contents it changes sign giving k = 0.019 for a Ge content of 40% and k = 0.131 for a Ge content of 20%.

In quantum dots, hole motion is confined in all directions. This results in the mixing of HH and LH states which can be strongly influenced by the additional presence of strain. It has been theoretically shown[15] that in the case of small, pure Ge self-assembled nanocrystals with pyramidal shape, confinement and compressive strain cause a significant HH-LH splitting.

The first confined state, closest to the valence-band edge, has a dominant HH character leading to a large $|g_\perp|$ (~12) and a pronounced anisotropy ($g_\perp/g_{//} \sim 6$).

The $|g|$ factors and anisotropies measured in the present work are smaller due to different reasons: 1) The dome-shaped self-assembled nanocrystals have a bigger size which implies smaller confinement and strain; 2) The investigated nanocrystals do not consist of pure Ge (for the growth conditions used the Ge content is known to be in 50-75% range[18]).

**Supplementary Figure 1**

**Superconductivity of Al electrodes**

**a,** $dI_{SD}/dV_{SD}$ versus $(B_{//}, V_{SD})$ for $V_G$ fixed at the position of the blue line in Fig. 5a. The onset of elastic (black arrow) and inelastic (yellow arrow) cotunneling can be observed. **b-c,** Schematic energy diagrams illustrating the onset of elastic cotunneling at $V_{SD} = 2\Delta$ and of inelastic cotunneling at $V_{SD} = 2\Delta + \delta$, respectively.

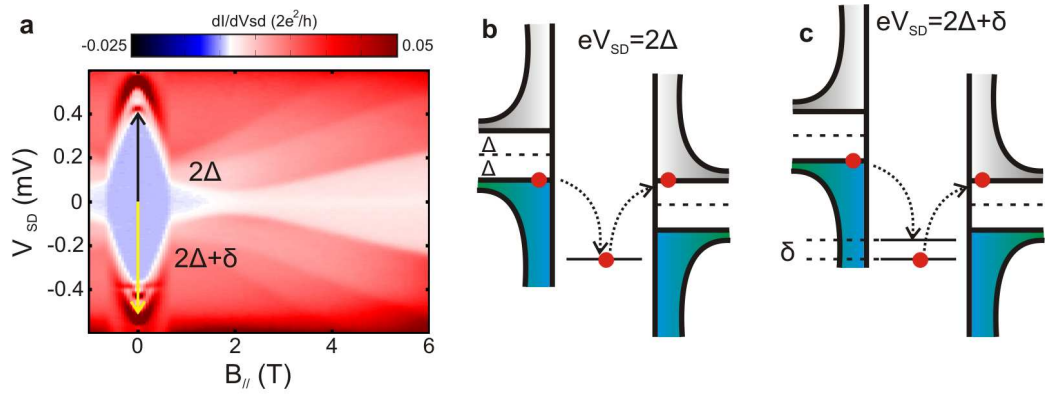

Fig. S1